\documentclass[aps,prd,amsmath,amssymb,twocolumn]{revtex4}

\usepackage{bm}
\usepackage{amsmath}
\usepackage{amssymb}
\usepackage{latexsym}
\usepackage{amsfonts}
\usepackage{epsfig}
\usepackage{psfrag}
\usepackage{graphicx}

\usepackage{graphicx}
\usepackage{mathptmx}      
\usepackage{latexsym}

\usepackage{amssymb}
\usepackage{amscd}
\usepackage{amsmath}
\usepackage{pstricks}
\usepackage{pst-all}
\usepackage{epsf}
\usepackage{latexsym}
\usepackage{fleqn}
\usepackage{psfrag}
\usepackage{epsfig}
\usepackage{bm}

\newcommand{\ord}{{\cal O}}
 
\newcommand{\ket}[1]{\left|#1\right>} 
\newcommand{\bra}[1]{\left<#1\right|} 
\newcommand{\braket}[2]
{\left<#1|#2\right>}

\newcommand{\nn}{\nonumber\\} 
\newcommand{\ul}{\underline}

\newcommand{\na}{\mbox{\boldmath$\nabla$}}
\newcommand{\bea}{\begin{eqnarray}}
\newcommand{\ea}{\end{eqnarray}}

\begin{document}

\title{Dynamical quantum phase transitions}

\author{Ralf Sch\"utzhold}

\email{ralf.schuetzhold@uni-due.de}

\affiliation{Institut f\"ur Theoretische Physik,
Technische Universit\"at Dresden, D-01062 Dresden, 
Germany
\\
Fachbereich Physik, Universit\"at Duisburg-Essen, 
D-47048 Duisburg, Germany
}



\begin{abstract}
A sweep through a quantum phase transition by means of a time-dependent 
external parameter (e.g., pressure) entails non-equilibrium phenomena 
associated with a break-down of adiabaticity: At the critical point, 
the energy gap vanishes and the response time diverges 
(in the thermodynamic limit). 
Consequently, the external time-dependence inevitably drives the system
out of equilibrium, i.e., away from the ground state, if we assume zero 
temperature initially. 
In this way, the initial quantum fluctuations can be drastically amplified 
and may become observable -- especially for symmetry-breaking (restoring)
transitions. 
By means of several examples, possible effects of these amplified quantum 
fluctuations are studied and universal features (such as freezing) are 
discussed.
%
%
%
%
\end{abstract}

\maketitle

\section{Motivation}\label{intro}

In many cases, the properties of the ground state (or, at finite temperatures, 
the thermal equilibrium state) are sufficient for understanding the behavior
of a given condensed-matter system.
In many other scenarios, however, the actual quantum state shows significant 
deviations from the ground state and non-equilibrium phenomena have to be
taken into account.
Even if the system starts near the (initial) ground state, it may depart 
during the evolution -- for example, due to an external time-dependence.

For example, let us consider an explicitly time-dependent Hamiltonian with
a discrete (non-degenerate) spectrum 
$H(t)\ket{\psi_n(t)}=E_n(t)\ket{\psi_n(t)}$.
After inserting this instantaneous eigenvector expansion into the 
Schr\"odinger equation and some approximations, one obtains the adiabatic 
expansion 
\bea
\label{adiabatic}
\ket{\psi(t)}
\approx
\ket{\psi_0(t)}
+
\sum_{n>0}
\frac{\bra{\psi_n}\dot{H}\ket{\psi_0}}{(E_n-E_0)^2}
\,e^{i\varphi_{n}}\,\ket{\psi_n(t)}
\,.
\ea
I.e., if the system started in its ground state $\ket{\psi_0(t_0)}$, 
its actual quantum state $\ket{\psi(t)}$ will remain near 
\cite{Born} the 
instantaneous ground state $\ket{\psi_0(t)}$ if the non-adiabatic 
corrections are small, i.e., if 
$\bra{\psi_n}\dot{H}\ket{\psi_0}\ll(E_n-E_0)^2$.
The latter condition compares the rapidity of the external time 
dependence $\dot{H}$ with the internal response times given by the
energy level spacings $\Delta E_n=E_n-E_0$.
If the response time of the system is short enough such that it can 
adapt to the external variation, it will stay near the ground state 
-- if the externally imposed change is too fast, however, the system 
cannot respond and will depart from the instantaneous ground state 
$\ket{\psi_0(t)}$, leading to the creation of excitations 
$\ket{\psi_{n>0}}$, e.g., quasi-particles. 

This observation directly indicates the close ties between 
(quantum) phase transitions \cite{Sachdev} 
and non-equilibrium phenomena:
At the critical point, at least some of the energy levels converge 
$\Delta E_n=(E_n-E_0)\downarrow0$ and hence the associated response 
times diverge $\Delta E\Delta t\sim\hbar$, which makes it very easy 
to depart from equilibrium, even for very slow external time 
dependences $\dot{H}$, see, e.g., \cite{Dorner,Dziarmaga,Damski} 
Of course, the resulting non-equilibrium phenomena do not just depend on 
the energy levels $\Delta E_n=(E_n-E_0)$, the matrix elements 
$\bra{\psi_n}\dot{H}\ket{\psi_0}$ are also very important. 

\section{First Example}\label{first}

\begin{figure}
\includegraphics[width=0.4\textwidth]{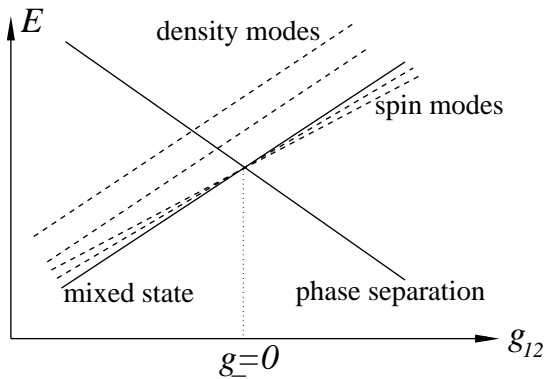}
\caption{Sketch of the level structure from Eq.~(\ref{two-component})
near the critical point $g_-=0$ separating the mixed condensate from 
the phase separated state.
The two solid lines denote these two competing ground states which cross 
at $g_-=0$ and the quasi-particle excitations are depicted by dashed lines.}
\label{level}        
\end{figure}

Let us start by considering a very simple example: a homogeneous 
two-component Bose-Einstein condensate \cite{two-component}. 
In the dilute-gas limit, the Hamiltonian density reads  
\bea
\label{two-component}
\hat{\cal H}
=\sum_{ab}
\hat{\Psi}^\dagger_a
\left(-\frac{\na^2}{2m}
+\frac{g_{ab}}{2}\hat{\Psi}^\dagger_b\hat{\Psi}_b
\right)\hat{\Psi}_a
\,,
\ea
where $\hat{\Psi}_a=(\hat{\Psi}_1,\hat{\Psi}_2)$ denote the field operators 
of the two components, $m$ the mass of the condensed particles, 
and $g_{ab}$ the coupling matrix in the $s$-scattering approximation. 
Using the mean-field approximation and Madelung spilt 
$\hat{\Psi}_a\approx{\Psi}_a=\sqrt{\varrho_a}\exp(i\Phi_a)$ with the 
densities $\varrho_a>0$ and the phases $\Phi_a$, we obtain two decoupled 
Goldstone modes due to the broken $U(1)\otimes U(1)$ symmetry 
$\Phi_\pm=(\delta\Phi_1\pm\delta\Phi_2)/\sqrt{2}$.
The hard density mode $\Phi_+$ corresponds to variations of the total
density $\varrho=\varrho_1+\varrho_2$ while the soft spin mode $\Phi_-$ 
is associated to the density difference $\varrho_1-\varrho_2$.
The propagation velocities of the two modes are given by the eigenvalues 
$g_\pm=(g_{11}+g_{22})^2/4\pm\sqrt{(g_{11}-g_{22})^2/4+g_{12}^2}$
of the coupling matrix $g_{ab}$. 
Depending on the relation between the inter-component repulsion 
$g_{12}=g_{21}>0$ and the intra-component couplings $g_{11},g_{22}>0$, 
there is a phase transition when $g_-$ changes its sign.
If the repulsion within the same component $g_{11},g_{22}$ dominates,
the condensate will be in the homogeneous mixed state 
$\varrho_1=\varrho_2$.
In case the particles of different components repel each other more 
strongly, the inhomogeneous phase separated state 
(either $\varrho_1=0$ or $\varrho_2=0$) will be energetically 
favorable, see Fig.~\ref{level}.

\begin{figure}
\includegraphics[width=0.4\textwidth]{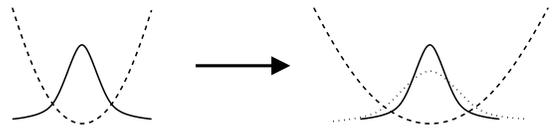}
\caption{Schematic of the squeezing (amplification) of the quantum 
fluctuations.
Each (linearized) quasi-particle mode corresponds to a harmonic oscillator,
whose potential is depicted as a dashed curve. 
Initially (left picture), the mode is in its ground state (solid curve).
Approaching the critical point, the modes become softer and hence the 
potential gets more shallow (right picture). 
As long as this happens slow enough [cf.~Eq.~(\ref{adiabatic})], 
the quantum state evolves adiabatically and stays near the 
instantaneous ground state (dotted curve). 
After freezing, however, the modes cannot adapt to this change anymore 
and their quantum state (solid curve in right picture) deviates from the 
ground state (dotted curve). 
In view of the linearity of the equations of motion, the quantum state 
evolves into a squeezed state (e.g., the wave function is still Gaussian).
This squeezing effect becomes even more pronounced if the quadratic 
potential turns over, as in Fig.~\ref{potentials}. 
The direction in which the quantum fluctuations will be amplified 
depends on the system under consideration.
For a symmetry-breaking transition 
(where the potential turns over, cf.~Fig.~\ref{second}), 
for example, the quantum fluctuations typically grow in the direction 
of symmetry-breaking.}
\label{squeezing}        
\end{figure}

Now, let us study the sweep through this phase transition by means of a 
time-dependent $g_{12}(t)$ starting in the mixed state.
The hard density modes $\Phi_+$ will basically remain unaffected by the 
transition, but the spin modes $\Phi_-$ become arbitrarily soft when 
approaching the critical point $g_-=0$ and finally become unstable on the
other side.
Hence these modes are the most interesting ones. 
In such a sweep, there will be four major stages:
\begin{itemize}
\item{\bf cooling}\quad 
Since approaching the critical point implies a reduction of the 
quasi-particle energies $\Delta E_n=E_n-E_0$ of the spin modes, their 
temperature (if non-zero initially) will drop as long as we are still 
approximately in equilibrium (i.e., adiabatic). 
\item{\bf freezing}\quad 
Close enough to the critical point, however, the spin modes become too soft,  
i.e., slow, and thus cannot adopt to the externally imposed variation 
$g_{12}(t)$ anymore, i.e., they freeze.  
\item{\bf squeezing}\quad 
Even if we started in the initial ground state, this freezing of the modes
implies a squeezing (and thus amplification) of the quantum fluctuations 
due to the breakdown of adiabaticity, cf.~Eq.~(\ref{adiabatic}) and 
Fig.~\ref{squeezing}.  
\item{\bf re-heating}\quad 
Finally, far enough on the other side of the phase transition, the system
starts to ``realize'' that it is not in its true ground state and the spin
modes star to grow exponentially -- with the frozen and amplified quantum 
fluctuations as initial seeds. 
\end{itemize}
Interestingly, this sequence: cooling $\to$ freezing $\to$ squeezing $\to$
re-heating exhibits striking similarities to cosmic inflation.
Indeed, one might speculate whether cosmic inflation was not a real 
expansion of the universe over many orders of magnitude, but instead
represent our distorted view of an early cosmic phase transition 
\cite{inflation/deflation}. 

\section{Horizon Analogue}\label{horizon}

Let us study the general behavior mentioned in the previous Section in 
some more detail.
To this end, we have to consider the equation of motion for the spin modes.
Of course, we could derive it directly from the original Hamiltonian 
(\ref{two-component}). 
However, for the sake of universality, let us base our discussion on 
general arguments instead.
Assuming that the spin modes can be described by a linearized low-energy 
effective action, we write down the most general form for scalar Goldstone
modes in a homogeneous and isotropic medium \cite{inflation/deflation} 
\bea
\label{low-energy}
{\cal L}_{\rm eff}=\frac{1}{2}
\left(\frac{1}{\alpha[g_-(t)]}\,\dot\Phi^2_-
-\beta[g_-(t)]\,(\na\Phi_-)^2\right)
\,,
\ea
where the two remaining factors $\alpha$ and $\beta$ are functions of the 
external parameter $g_-(t)$ and have to be determined from the underlying 
system.
Accordingly, the effective energy (Hamiltonian) density of these 
Goldstone modes reads 
\bea
{\cal H}_{\rm eff}=\frac{1}{2}
\left(\alpha[g(t)]\,\Pi^2+\beta[g(t)]\,(\na\Phi)^2\right)
\,.
\ea
Before the transition, the spin modes are stable and hence $\alpha$ and 
$\beta$ must be positive.
After the transition, however, the spin modes become unstable and thus 
at least one of the two parameters has to change its sign, i.e., it must 
vanish at the critical point. 
As a result, the propagation velocity $c^2=\alpha\beta$ must also go to 
zero when approaching the transition, which precisely reflects the fact 
that the spin modes become infinitely soft. 

The consequence of such an evanescent propagation velocity $c(t)$ 
can nicely be explained by the analogy to gravity/cosmology:
For a constant speed $dc/dt=0$, quasi-particle excitations may 
propagate arbitrarily far through the sample 
(in the absence of damping etc.) if we wait long enough. 
If $c(t)$ decreases sufficiently fast, however, the excitations may not 
travel infinitely far -- even if one waits long enough and takes an 
infinite time to reach the critical point -- but merely a finite 
distance, which is an exact analogue of a cosmic horizon 
%
\bea
\label{horizon-size}
\Delta r(t)=\int\limits^{\infty}_{t}dt'\,c(t')
\,.
\ea
The horizon size $\Delta r(t)$ corresponds to the maximum distance 
covered by quasi-particles starting at time $t$ and hence measures 
the range of causal connection \cite{horizons}. 
Of course, in a real experiment, the critical point is crossed after 
a finite time, which should then be used as the upper limit
(with the physics conclusions remaining unchanged). 

Two points at distances larger than the horizon $\Delta r(t)$ can no 
longer exchange information or energy, i.e., the horizon describes a 
dynamical loss of causal connection.
Since maintaining equilibrium requires causal connections between all
points of the sample (e.g., in order to equilibrate local density 
variations), this loss of causal connection indicates the departure from 
equilibrium and the breakdown of adiabaticity.
As mentioned before, this typically entails an amplification of the quantum 
fluctuations, which can be understood in the following way:
Since the analogue horizon shrinks steadily 
\bea
\frac{d}{dt}\Delta r(t)
=
\frac{d}{dt}\int\limits^{\infty}_{t}dt'\,c(t')
=
-c(t)
\,,
\ea
it will engulf every wavelength $\lambda$ after a given time.
Therefore, all modes $\lambda$ will pass through the three main stages 
of their evolution:
\begin{itemize}
\item{\bf oscillation}\quad
Initially $\lambda\ll\Delta r(t)$, the modes do not notice the horizon 
and oscillate almost freely (though with a decreasing frequency, 
cf.~the cooling in the previous Section). 
\item{\bf horizon crossing}\quad
At some point of time, the horizon closes in $\lambda\sim\Delta r(t)$ 
and inhibits further oscillations because crest and trough of a wave 
are separated by the horizon and hence cannot exchange energy anymore. 
\item{\bf freezing \& squeezing}\quad
After that, the modes are frozen and cannot respond to the external 
change anymore, i.e., the initial Gaussian ground state will turn into
a squeezed state representing the amplification of the quantum 
fluctuations, cf.~Fig.~\ref{squeezing}.  
\end{itemize}
According to our standard model of cosmology, precisely the same mechanism
occurred in the early universe (during inflation) and is responsible for the 
generation of the seeds for structure formation out of the initial quantum 
vacuum fluctuations. 
Traces of these amplified quantum vacuum fluctuations can still be observed 
today in the anisotropies of the cosmic microwave background radiation. 

\section{Universality?}\label{universal}

\begin{figure}
\includegraphics[width=0.4\textwidth]{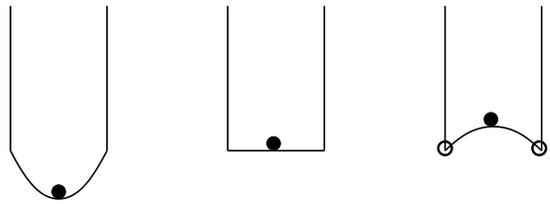}
\caption{Sketch of the energy landscape associated to
Eq.~(\ref{two-component}). 
The filled black dot denotes the initial ground state (mixed condensate) 
and the hollow dots correspond to the final ground states (phase separation).
In view of the jump in the ground state structure, the phase transition is
of first order -- though it is a somewhat untypical example, since the energy
landscape becomes flat (in the relevant direction) at the critical point 
(middle panel).}
\label{first-irreg}       
\end{figure}

Now, after having studied this specific example, one is lead to the question
of how universal these phenomena really are. 
In order to study this point, let us have a look at the effective energy 
landscape in terms of the order parameter $x=\varrho_1/(\varrho_1+\varrho_2)$.
It is restricted according to $0\leq x\leq1$ with the boundary values 
representing the phase separated region while the mixed state is given by 
$x=1/2$.
Inserting this ansatz into the Hamiltonian (\ref{two-component}), the 
relevant part of the energy functional reads $g_-(x-1/2)^2$, which is 
plotted in Fig.~\ref{first-irreg}.
Even though the phase transition is of first order (on the mean-field level),
it is a somewhat untypical example, because the energy landscape becomes
flat at the critical point. 
Therefore, adding a small non-linearity such as $(x-1/2)^4$ would turn it 
into a transition of second order.

\begin{figure}
\includegraphics[width=0.4\textwidth]{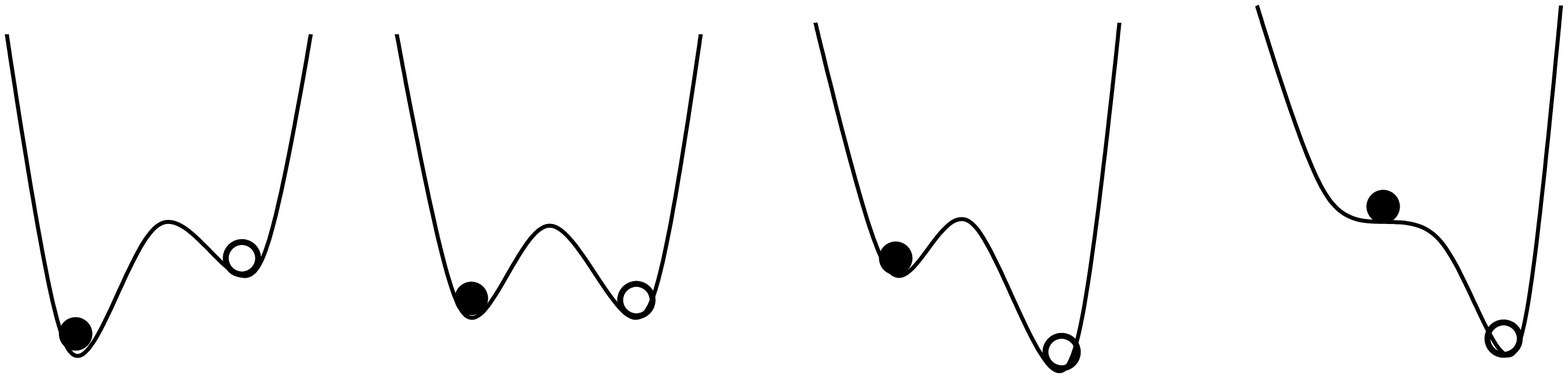}
\caption{Sketch of the energy landscape associated to a typical first-order
phase transition. 
Again the filled/hollow black dot denotes the initial/final ground state. 
In this case, the vicinity of the initial ground state remains (locally) 
stable at
the critical point (second panel from the left) and the linearized 
quasi-particle modes do not become arbitrarily soft at this stage. 
Typically, this happens much later, at the point of instability of the 
initial ground state (right panel).}
\label{first-reg}       
\end{figure}

In contrast, the energy landscape of a typical first-order transition is 
sketched in Fig.~\ref{first-reg}.
In this case, the vicinity of the initial ground state remains stable even 
after the transition and there is no local linearized mode connecting the 
two competing vacua before one reaches the point of instability, 
cf.~Fig.~\ref{first-reg}, right panel. 
Therefore, the previous arguments (softening of modes etc.) do not apply 
to the critical point, but they could be applied to the point of 
instability (which occurs later).

\begin{figure}
\includegraphics[width=0.4\textwidth]{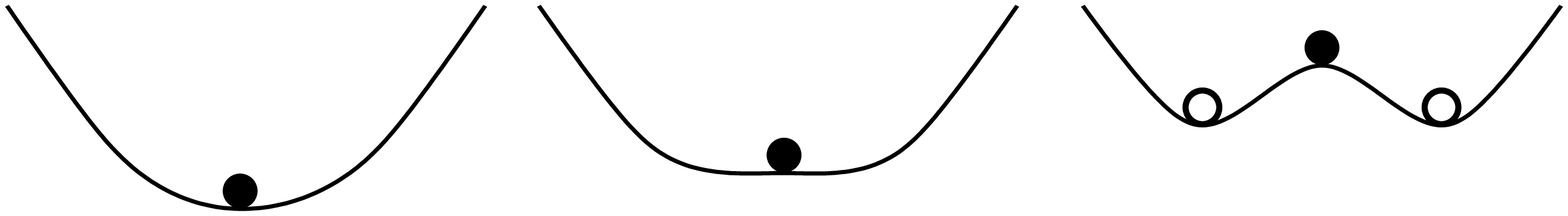}
\caption{Sketch of the energy landscape associated to a typical 
symmetry-breaking second-order transition. 
At the critical point (middle panel), the local curvature in the vicinity 
of the initial ground state (filled dot) changes its sign and hence the 
associated quasi-particle modes become arbitrarily soft at the transition 
and turn unstable afterwards. 
The time-reversed sweep corresponds to a symmetry-restoring transition. 
In this case, the ground state (hollow dot) does not become unstable, 
but still the associated quasi-particle modes become arbitrarily soft 
at the transition.}
\label{second}       
\end{figure}

The aforementioned instability of the transition in Fig.~\ref{first-irreg} 
towards the transformation into a second-order transition already suggests
that the phenomena discussed in the previous Sections could also be relevant 
for transitions of higher order.
Fig.~\ref{second} depicts the energy landscape of a typical symmetry-breaking
second-order phase transition. 
As suggested by this picture, there will be some modes which become 
arbitrarily soft at the critical point and unstable afterwards, i.e., 
during the sweep through such a transition, one generally reproduces 
the sequence: cooling $\to$ freezing $\to$ squeezing $\to$ re-heating. 
Transversing the critical point in the opposite direction 
(i.e., a symmetry-restoring transition), there are still modes 
which become arbitrarily soft -- but they do not necessarily become 
unstable.
Consequently, the shorter sequence: cooling $\to$ freezing $\to$ squeezing
does also apply here, but the re-heating part requires further 
considerations. 

\begin{figure}
\includegraphics[width=0.4\textwidth]{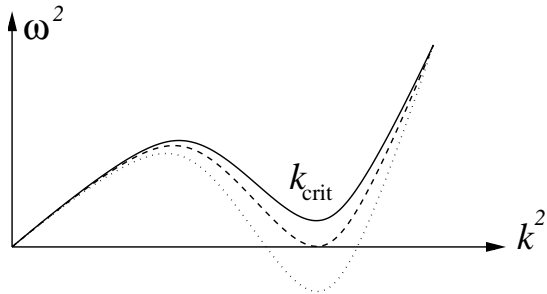}
\caption{Typical evolution of the dispersion relation where the instability 
occurs at a finite wavenumber $k_{\rm crit}$. Before the transition
(solid curve), the frequencies are non-negative. 
At the critical point (dashed curve), the ``roton'' dip touches the axis 
and afterwards, the modes in the vicinity of $k_{\rm crit}$ start to grow 
(dotted curve).}
\label{disp-roton}       
\end{figure}

\begin{figure}
\includegraphics[width=0.4\textwidth]{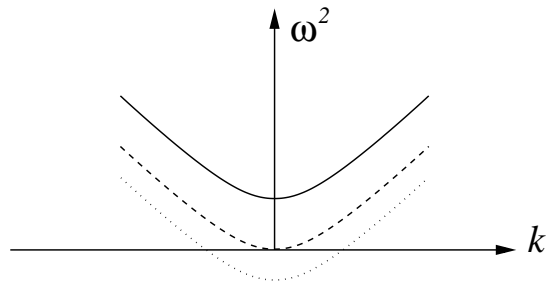}
\caption{Typical evolution of the dispersion relation where the instability 
occurs at $k=0$ since the quasi-particle gap -- i.e., their effective mass 
$m^2$ in Eq.~(\ref{mass-gap}) -- changes sign. 
}
\label{disp-mass}       
\end{figure}

\begin{figure}
\includegraphics[width=0.4\textwidth]{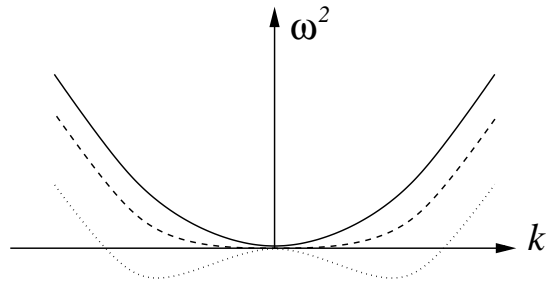}
\caption{Typical evolution of the dispersion relation where the instability 
occurs at $k=0$ since the quasi-particle stiffness $c^2$ of the Goldstone 
modes changes sign, cf.~(\ref{low-energy}).}
\label{disp-speed}       
\end{figure}

Of course, having established that some modes become arbitrarily soft when
approaching the critical point does not mean that they behave like the spin 
modes in Eq.~(\ref{low-energy}). 
For the example in Sec.~\ref{first}, the typical length scale of the 
amplified quantum fluctuations was set by the horizon size and thus the 
sweep rate $dc/dt$.  
This is not always the case.
In order to explore the different possibilities, let us study 
quasi-particle modes which become arbitrarily soft when approaching the 
critical point (and unstable afterwards).
Their relevant features should be encoded in the quasi-particle 
dispersion relation $\omega({\vec k})$.
Assuming invariance under time-reversal and reflection as well as (spatial) 
rotations\footnote{A counter-example would be a phase gradient and thus 
flow through our sample, cf.~Sec.~\ref{extended}}, it simplifies to 
$\omega^2({\vec k}^2)$.
The critical point then occurs when $\omega^2({\vec k}^2)$ starts to dive 
below the vertical axis. 
This could happen at non-zero $k$, see Fig.~\ref{disp-roton}, 
or at zero $k$.
In the latter case, we may employ a Taylor (i.e., low-energy) 
expansion similar to Eq.~(\ref{low-energy}) 
\bea
\label{mass-gap}
\omega^2({\vec k}^2)=m^2+c^2{\vec k}^2+\ord({\vec k}^4)
\,.
\ea
Thus there are two options for an instability: either $m^2$ changes its
sign, see Fig.~\ref{disp-mass}, or $c^2$ changes its sign while $m=0$, 
see Fig.~\ref{disp-speed}. 
Note that the case of $c^2$ changing its sign in the presence of a 
positive $m^2$ would bring us back to the case in Fig.~\ref{disp-roton}. 

An example for the scenario in Fig.~\ref{disp-roton} is the extended 
Bose-Hubbard model (cf.~Sec.~\ref{extended}) where the dispersion relation 
may possess a ``roton'' dip which dives below the vertical axis at the 
superfluid-supersolid phase transition. 
In this case, the typical length scale of the amplified quantum fluctuations
is naturally set by the characteristic wavenumber $k_{\rm crit}$ at the 
``roton'' dip, which determines the period of the supersolid order and is 
independent of the sweep rate etc. 
However, spatial modulations around this wavenumber $k_{\rm crit}$ 
(which could induce defects in the supersolid crystal, for example) 
may depend on the sweep rate.

The instability at small $k$ sketched in Fig.~\ref{disp-mass} occurs
in the example discussed in Sec.~\ref{spinor}. 
In contrast to Fig.~\ref{disp-roton}, the characteristic length 
scale cannot be read off the critical dispersion relation 
(dashed curve) in general. 
Instead, it is set by the sweep rate and/or the final 
state\footnote{The initial state should not be important as the system 
is supposed to be adiabatic initially.}, depending on the system 
parameters. 
For the experiment \cite{Sadler} discussed in Sec.~\ref{spinor}, 
the final state determines the characteristic length scale.
In other regions of parameter space (e.g., with a much slower quench), 
however, the sweep rate may yield the dominant contribution. 
In this situation, a quantum version of the Kibble-Zurek 
\cite{Kibble,Zurek} scaling arguments \cite{Zurek-scaling} 
may be applied:  
The time-dependent correlation length $\xi(t)\propto1/m(t)$ is cut off 
at the time $\tilde t$ when adiabaticity breaks down, i.e.,  
$\tilde t\simeq t_{\rm response}(\tilde t)\propto1/\Delta E(\tilde t)$.
This saturated correlation length $\tilde\xi=\xi(\tilde t)$ then 
determines the typical distance of the topological defects. 

Finally, the case of Goldstone modes (which become unstable upon 
crossing the transition) is sketched in Fig.~\ref{disp-speed}.
For example, the system discussed in Sec.~\ref{first} falls into 
this category.
Note that Fig.~\ref{first-irreg} describes the behavior of that system 
under large but homogeneous (i.e., $k=0$) displacements, whereas the 
dispersion relation in Fig.~\ref{disp-speed} corresponds to small 
(linearized) inhomogeneous displacements with various $k$.  
In this case, the concept of a horizon analogue sketched in 
Sec.~\ref{horizon} can be applied.
In the vicinity of the critical point (dashed line in Fig.~\ref{disp-speed}), 
the time-dependent horizon size in Eq.~(\ref{horizon-size}) determines the 
spectrum \cite{inflation/deflation} and the typical length scales of the 
amplified quantum fluctuations, but after the transition 
(dotted line in Fig.~\ref{disp-speed}), the final state typically induces 
an additional length scale corresponding to the fastest growth of the 
modes (negative minima of dispersion curve). 
Note that the Kibble-Zurek scaling arguments sketched above do not apply 
in this situation:
Modes with different $k$ cross the horizon in Eq.~(\ref{horizon-size}) 
and thus freeze (i.e., become non-adiabatic) at different times -- 
hence, there is no unique time $\tilde t$ and consequently no 
saturated correlation length $\tilde\xi=\xi(\tilde t)$. 

\section{Bose-Hubbard Model}\label{bose-hubbard}

Now, after having discussed some general aspects, let us turn to a few 
explicit examples.  
A prototypical example \cite{Sachdev} for a second-order phase transition 
is the Bose-Hubbard model 
\bea
\label{Bose-Hubbard-ham}
\hat H
=
J(t)\sum\limits_{\alpha\beta}
M_{\alpha\beta}\hat a_\alpha^\dagger\hat a_\beta
+\frac{U}{2}\sum\limits_\alpha(\hat a_\alpha^\dagger)^2\hat a_\alpha^2
\,. 
\ea
As usual, $\hat a_\alpha^\dagger,\hat a_\beta$ are the (bosonic) 
creation/annihilation operators for the lattice sites $\alpha$ and 
$\beta$, respectively. 
Hence, the first term describes hopping between the lattice sites 
$\alpha$ and $\beta$, where the lattice structure (e.g., cubic) is 
encoded in the matrix $M_{\alpha\beta}$ and the tunneling rate $J(t)$ 
is chosen to be time-dependent. 
The second term describes the fact that two bosons repel each other 
if they are on the same lattice site, where $U$ denotes the associated 
energy penalty.  
For integer filling 
$n=\langle\hat a_\alpha^\dagger\hat a_\alpha\rangle
=\langle\hat n_\alpha\rangle\in\mathbb N$, 
the system undergoes a second-order quantum phase 
transition\footnote{Strictly speaking, there is no Bose-Einstein 
condensation and thus also no phase transition in a one-dimensional 
homogeneous lattice of infinite length.
Therefore, we should consider either lattices of two or more spatial 
dimensions or chains of sufficiently small length. 
In the latter case, however, there is no sharp phase transition but 
only an approximate one (cf.~Sec.\ref{finite}).}
at a critical value of the hopping rate $J_{\rm crit}=\ord(U/n)$.
For $J \gg U/n$, the first term dominates and the bosons can move 
freely on the lattice, which corresponds to the superfluid phase 
\bea
\label{coherent}
J \gg U/n
\,\leadsto\,
\ket{\Psi}_{\rm sf}\propto
\left(\sum_{\alpha}\hat a_\alpha^\dagger\right)^N
\ket{0}
\,,
\ea
where $N=\sum_{\alpha}n$ is the total number of bosons on the lattice. 
In the opposite limit $J \ll U/n$, the second term dominates and causes 
the particles to lock in the lattice sites, i.e., they are not free 
to move across the lattice anymore.
Hence this phase is called the Mott insulator state 
\bea
\label{Mott}
J \ll U/n
\,\leadsto\,
\ket{\Psi}_{\rm Mott}
\propto\bigotimes\limits_{\alpha}(\hat a_\alpha^\dagger)^n
\ket{0}
\,.
\ea
In the Mott insulator phase, the ground state is separated by a finite 
gap from the other excited states.
The superfluid state, on the other hand, is associated to a certain 
phase and thus breaks the $U(1)$ symmetry 
$\hat a_\alpha\to e^{i\varphi}\hat a_\alpha$
of the Hamiltonian (\ref{Bose-Hubbard-ham}).
As a result, this state supports gap-less Goldstone modes, which are the 
lattice phonons. 

Now, let us consider the following scenario: starting deep in the 
superfluid phase $J \gg U/n$, we decrease the hopping rate $J(t)$ and
thereby sweep through the superfluid-Mott phase transition far into the 
Mott regime $J \ll U/n$.
If we did this infinitely slowly, we would end up in the Mott state
(\ref{Mott}). 
A very rapid sweep, on the other hand, would leave the system no time 
to respond and thus it would stay in the initial coherent state 
(\ref{coherent}). 
For a finite sweep velocity $dJ/dt$, one would expect to end up 
somewhere in between the two states. 
A good marker for distinguishing the two regimes is the (final) 
number variance $\Delta^2(n_\alpha)=\langle\hat n_\alpha^2\rangle-
\langle\hat n_\alpha\rangle^2$.
The coherent state generates Poissonian number statistics and 
these fluctuations are large $\Delta^2(n_\alpha)=n$ in the superfluid 
state $J \gg U/n$, whereas they vanish deep in the Mott 
state~(\ref{Mott}).  

Because the Hamiltonian (\ref{Bose-Hubbard-ham}) cannot be diagonalized 
analytically, some approximations are necessary for deriving explicit 
results \cite{Bose-Hubbard-short,Bose-Hubbard-long}. 
Here, we assume a large filling $n\gg1$, which facilitates an expansion 
into (inverse) powers of $n$ via 
$\hat n_\alpha=n+\delta\hat n_\alpha+\ord(n^0)$ where the fluctuations 
$\delta\hat n_\alpha=\ord(\sqrt{n})$ can be treated analytically. 
These fluctuations correspond to the Goldstone modes mentioned earlier 
and their propagation speed (in the continuum limit) can be estimated via 
$c^2=\ell^2JUn$, where $\ell$ is the lattice spacing. 
Thus, if $J$ decreases fast enough, we get a horizon analogue as discussed 
in Sec.~\ref{horizon}. 

Let us study a few examples for the temporal behavior of the sweep:
For an exponentially decaying hopping rate $J(t)=J_0\exp\{-\gamma t\}$,
a horizon emerges for all values of $\gamma$ and hence the Goldstone 
modes freeze, cf.~Sec.~\ref{horizon}.
The frozen number fluctuations are given by 
\bea
\Delta^2(n_\alpha)
=
\langle\hat n_\alpha^2\rangle-
\langle\hat n_\alpha\rangle^2
=
\langle\delta\hat n_\alpha^2\rangle
=
n\frac{1-e^{-2\pi\nu}}{2\pi\nu}
\,,
\ea
where the adiabaticity parameter $\nu=Un/\gamma=\mu/\gamma$ is given by 
the ratio of the internal time scale (chemical potential $\mu=Un$) over
the external sweep rate $\gamma$ and thus measures the rapidity of the 
sweep.
In agreement with the previous arguments, we recover the limiting cases 
$\Delta^2(n_\alpha)\uparrow n$ for $\nu\downarrow0$ and 
$\Delta^2(n_\alpha)\downarrow0$ for $\nu\uparrow\infty$.
If the hopping rate merely exhibits a slow polynomial decay 
$J(t)\propto t^{-x}$ with $x<2$, no horizon emerges and the number
fluctuations do not freeze at a finite time but oscillate forever.
The boundary case $J(t)\propto t^{-2}$ is quite interesting: 
Since it marks the limit for horizon formation, the modes show a 
behavior in between freezing at a finite value and eternal oscillation, 
i.e., they decay down to zero quite rapidly.
This situation is analogous to a damped harmonic oscillator, where 
the critical damping [corresponding to $J(t)\propto t^{-2}$] marks
the border between the over-damped (horizon formation and freezing)
and the under-damped case (eternal oscillation). 
Hence this sweep dynamics $J(t)\propto t^{-2}$ would be the optimal 
choice for approaching the Mott state with 
$\Delta^2(n_\alpha)\downarrow0$ most quickly
\cite{Bose-Hubbard-short,Bose-Hubbard-long}. 

\section{Extended Bose-Hubbard Model}\label{extended}

If we extend the Hamiltonian (\ref{Bose-Hubbard-ham}) and include 
inter-site interactions given by $V_{\alpha\beta}$
\bea
\label{Bose-Hubbard-ext}
\hat H
=
\sum\limits_{\alpha\beta}
\left(
T_{\alpha\beta}\hat a_\alpha^\dagger\hat a_\beta
+V_{\alpha\beta}\hat n_\alpha\hat n_\beta
\right)
\,, 
\ea
the phase diagram becomes much richer and now includes supersolid states etc. 
The superfluid-supersolid phase transition is a nice example for the 
scenario in Fig.~\ref{disp-roton} where the dispersion relation develops
a ``roton'' dip which touches the vertical axis at the critical point
(at zero temperature, cf.~\cite{supersolid}). 
This implies an instability towards density modulations at a critical 
wavenumber $k_{\rm crit}$, which corresponds to the period of the supersolid
order parameter. 
As before, the modes become arbitrarily soft when approaching the transition 
-- which entails interesting non-equilibrium effects.
For example, in the presence of an external phase gradient
(i.e., condensate flow), the branch of the ``roton'' dip corresponding to 
quasi-particles propagating against the condensate flow is closer to the 
axis, i.e., softer, than the other branch. 
Hence, approaching the transition too fast will create non-adiabatic 
excitations, which are stronger for these (softer) modes, because their 
gap is smaller, cf.~Eq.~(\ref{adiabatic}). 
As a result, non-equilibrium phenomena may induce a current 
(stemming from the amplified quantum fluctuations) in the 
{\em opposite} direction -- which might even compensate the condensate 
flow \cite{supersolid}. 
 
\section{Spinor ($S=1$) Condensate}\label{spinor}

As an example for the behavior sketched in Figs.~\ref{second} and 
\ref{disp-mass}, let us consider a three-component Bose-Einstein condensate, 
which can be represented by an effective spinor field operator
$\hat{\ul\Psi}=(\hat\Psi_x,\hat\Psi_y,\hat\Psi_z)$ and the Hamiltonian 
density \cite{Vortex-quantum}
\bea
\label{s=1}
\hat{\cal H}
&=&
\hat{\ul\Psi}^\dagger\cdot
\left(-\frac{\na^2}{2m}
+\frac{c_0}{2}\hat{\ul\Psi}^\dagger\cdot\hat{\ul\Psi}
\right)\hat{\ul\Psi}
\nn
&&
-\frac{c_2}{2}(\hat{\ul\Psi}^\dagger\times\hat{\ul\Psi})^2 
-q\hat\Psi_z^\dagger\hat\Psi_z
\,.
\ea
The first term is the same as in the case of one or two components 
(\ref{two-component}) and contains the pseudo-spin independent kinetic 
term $\na^2/(2m)$ and coupling $c_0$.
The second term $c_2$ corresponds to spin-1 scattering channel and 
favors the ferromagnetic phase with maximum magnetization
$\ul{F}=i\langle\hat{\ul\Psi}^\dagger\times\hat{\ul\Psi}\rangle\neq0$ 
due to $c_2<0$.
Finally, the third term $q>0$ denotes the quadratic Zeeman shift and 
would be minimized by placing all particles in the $z$-component, 
i.e., it favors the paramagnetic state 
$\langle\hat{\ul\Psi}^\dagger\times\hat{\ul\Psi}\rangle=0$. 
The (second-order) phase transition separating the two regimes occurs at
the critical point $q_{\rm crit}=2|c_2|\varrho$. 

\begin{figure}
\includegraphics[width=0.4\textwidth]{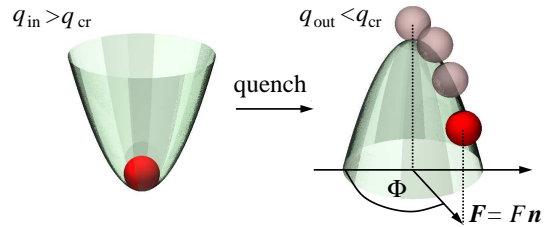}
\caption{Sketch of the energy landscape associated to Eq.~(\ref{s=1}). 
Initially, the system stays at $\ul{F}=0$ (paramagnetic phase).
After the transition, it starts to roll down the potential hill in some
direction [$O(2)$ symmetry] which is set by the initial quantum (or thermal)
fluctuations.}
\label{potentials}       
\end{figure}

Motivated by a recent experiment \cite{Sadler}, we study the sweep from 
the paramagnetic to the ferromagnetic phase, cf.~Fig.~\ref{potentials}.    
Initially, the system stays in the potential minimum at zero magnetization, 
but after the quench, the initial minimum at $\ul{F}=0$ turns to a local 
maximum and the system starts to roll down the potential hill, 
cf.~Fig.~\ref{potentials}.
Since this is a symmetry-breaking phase transition, the direction of 
descent can be arbitrary and is set by the initial quantum (or thermal) 
fluctuations. 
Thus the spontaneously formed (measurable) directions of the magnetization 
$\ul{F}$ provide a direct indicator for the initial quantum fluctuations, 
which were strongly amplified in that process. 
One result of these amplified quantum fluctuations could be the creation
of topological defects (which corresponds to the re-heating process 
mentioned in Sec.~\ref{first}) in a quantum analogue of the Kibble-Zurek 
\cite{Kibble,Zurek} mechanism, see also \cite{Dorner,Dziarmaga,Damski}.  

Since the $O(2)$ invariance of the Hamiltonian (\ref{s=1}) with respect to
rotations in the $x,y$-plane is broken by the ferromagnetic state, this 
phase supports topological (point) defects in the form of spin vortices 
in two spatial 
dimensions\footnote{In three spatial dimensions, point defects 
would not be possible for a broken $O(2)\simeq S_1$, but only line defects.
For a broken $SO(3)$, on the other hand, point defects would exist in 
three spatial dimensions due to $SO(3)/O(2)\simeq S_2$.}.
Such a defect corresponds to a spin configuration which cannot be deformed 
to a constant magnetization in a smooth way: 
Going around a vortex in a circle in real space, the magnetization 
describes a circle in the internal space of the ferromagnetic 
ground-state manifold, i.e., the magnetization always points towards 
the defect (vortex) or away from it (anti-vortex). 
The number of vortices (minus the number of anti-vortices) within a given 
area is the winding number $\mathfrak N$.
Calculating this winding number for a circle of radius $R$, we obtain 
(for large $R$) the scaling law \cite{Vortex-quantum} 
\bea
\langle\hat{\mathfrak N}^2\rangle\sim R\ln R
\,,
\ea
which lies in between the random phase walk model 
(predicting $\langle\hat{\mathfrak N}^2\rangle\sim R$) 
and the random vortex gas model 
(predicting $\langle\hat{\mathfrak N}^2\rangle\sim R^2$)
frequently discussed in the literature.  


\section{Finite-Size Scaling}\label{finite}

So far, our studies were devoted to the thermodynamic limit of infinitely 
large systems (cf.~the footnote in Sec.~\ref{bose-hubbard}). 
Strictly speaking, phase transitions are well-defined in this limit only. 
For systems of finite size, the sharp transition at a precisely localizable
critical point is typically broadened. 
Similarly, exact level crossings in the thermodynamic limit 
(as in Fig.~\ref{level}) usually become avoided level crossings for finite 
systems since boundary terms and finite-size effects lift the degeneracy 
in general.
As a result, the response time may not diverge at the critical point 
but remain bounded (even though very large).
In order to study non-equilibrium phenomena, the scaling of the response 
time (or the energy gaps) with system size is very important. 

As it turns out, this scaling may crucially depend on the order of the 
phase transition:
For a typical first order transition (cf.~Fig.~\ref{first-reg}), 
we have two competing ground states $\ket{\psi_<}$ and $\ket{\psi_>}$
which are locally distinguishable and become energetically favorable 
just before $\ket{\psi_<}=\ket{\psi_0(g<g_{\rm crit})}$ 
and after $\ket{\psi_>}=\ket{\psi_0(g>g_{\rm crit})}$ the transition. 
Due to their local distinguishability, their overlap decreases 
exponentially $\braket{\psi_<}{\psi_>}\sim\exp\{-\ord(n)\}$
with system size $n$ (e.g., number of lattice sites or spins).
In the low-energy sub-space spanned by $\ket{\psi_<}$ and $\ket{\psi_>}$,  
we have four relevant matrix elements of the Hamiltonian. 
At the critical point $g=g_{\rm crit}$, the diagonal elements coincide 
$\bra{\psi_<}H\ket{\psi_<}=\bra{\psi_>}H\ket{\psi_>}$, 
but the off-diagonal matrix elements 
$\bra{\psi_<}H\ket{\psi_>}=(\bra{\psi_>}H\ket{\psi_<})^*$ 
lead to an avoided level crossing with the fundamental gap 
$E_1-E_0=|\bra{\psi_<}H\ket{\psi_>}|$.
For a local Hamiltonian containing interactions of a limited number 
of sites (e.g., nearest neighbors) only $H\sim{\rm poly}(n)$,
the off-diagonal matrix elements scale exponentially 
$\bra{\psi_<}H\ket{\psi_>}\sim\exp\{-\ord(n)\}$ and hence the same 
scaling applies to the fundamental gap
$E_1-E_0\sim\exp\{-\ord(n)\}$.
This result nicely matches the intuitive picture based on the 
energy landscape in Fig.~\ref{first-reg}.
In order to stay in the ground state, the system has to tunnel through 
the energy barrier at the critical point. 
Since one would expect the height of the potential barrier to scale 
linearly with system size, the tunneling (i.e., response) time 
$\sim1/(E_1-E_0)$ should increase exponentially \cite{1st-vs-2nd}. 

For transitions of second or higher order, on the other hand, such a 
tunnelling barrier is absent and there are no locally distinct competing 
ground states at the critical point. 
Hence there is no reason for an exponential scaling of the response time
in this situation.
Indeed, for several analytically solvable examples such as the quantum 
Ising model in a transverse field, the response time is found to scale 
polynomially (instead of exponentially) with system 
size\footnote{However, the absence of the sketched reason for an 
exponential scaling does not necessarily imply that the minimum 
fundamental gap $(E_1-E_0)$ must scale polynomially for all second-order
transitions (as a counter-example, consider the random Ising chain).}. 
Furthermore, the ground state changes less abruptly in second-order
transitions than in those of first order, which also suggests that it 
should be easier to stay in the ground state. 

\section{Adiabatic Quantum Algorithms}\label{algorithms}

The intuition developed by considering the physics examples above might 
also be useful for gaining a deeper understanding in a rather different
subject -- quantum algorithms.  
Since the pioneering work of Shor and others, 
see, e.g., \cite{Shor,Grover}, it is known that quantum 
computers (which operate in the full Hilbert space of all quantum states)
should be able to solve certain problems much faster than all (known) 
classical algorithms. 
However, actually constructing a quantum computer of sufficient size is 
extremely hard since the fragile superpositions necessary for operating 
the quantum algorithm are very vulnerable to decoherence caused by the 
inevitable coupling to the environment. 
There are several ideas to overcome this problem: quantum error 
correcting codes, measurement-based (``one-way'') quantum computers, 
noise-resistant (e.g., topological) qubit realizations etc. 
In the following, we shall consider an alternative idea: adiabatic quantum 
algorithms \cite{Farhi}. 
In this scheme, the solution to a given problem is encoded into the 
ground state of an appropriately constructed problem Hamiltonian 
$H_{\rm out}$.
In order to arrive at the desired ground state of $H_{\rm out}$, one
starts with an initial Hamiltonian $H_{\rm in}$, whose ground state is 
known and can be prepared easily, and then slowly transform it to the 
final problem Hamiltonian $H_{\rm out}$, e.g., via a linear 
interpolation 
\bea
H(t)=[1-g(t)]H_{\rm in}+g(t)H_{\rm out}
\,.
\ea
If this interpolation is slow enough, we end up in (or near) the 
final ground state encoding the solution to our problem
\cite{estimate}. 
The computational complexity manifests itself in the runtime $T$, 
which is bounded from below by the response time determined by the 
minimum gap.
Now, for non-trivial problem Hamiltonians $H_{\rm out}$, there is 
typically a critical point where the fundamental gap $E_1-E_0$ between
the ground state and the first excited state becomes very small -- 
which bears strong similarities to a quantum phase transition, see also 
\cite{Latorre}.  

\begin{figure}
\includegraphics[width=0.4\textwidth]{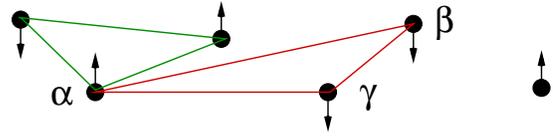}
\caption{Interaction topology of the Hamitonian (\ref{triples}) where each
triple $(\alpha,\beta,\gamma)$ corresponds to a triangle whose bonds govern 
the anti-ferromagnetic interaction $M_{\alpha,\beta}$.}
\label{clauses}       
\end{figure}

Let us study an explicit example. 
A rather interesting class of problems is \mbox{\em exact cover-3}, 
which corresponds to the following requirement:
Given a set of triples $(\alpha,\beta,\gamma)\in{\mathbb N}^3$, 
find at least one bit-string $z_\alpha\in\{0,1\}$ which satisfies 
the constraint $z_\alpha+z_\beta+z_\gamma=1$ for all triples 
$(\alpha,\beta,\gamma)$.
It can be shown that other tasks such as factoring can be mapped 
onto this problem class. 
A suitable problem Hamiltonian is given by the sum of 
${(\sigma^z_\alpha+\sigma^z_\beta+\sigma^z_\gamma+1)^2}$ 
over all these triples, which yields 
\bea
\label{triples}
H_{\rm out}
=
\sum\limits_{\alpha,\beta=1}^{n}
M_{\alpha,\beta}\sigma^z_\alpha\sigma^z_\beta
+
2\sum\limits_{\alpha=1}^{n}N_\alpha\sigma^z_\alpha
+\rm const.
\ea
This corresponds to a frustrated anti-ferromagnet in an external field
$N_\alpha$, where the interaction topology $M_{\alpha,\beta}$ is 
determined by the set of triples, cf.~Fig.~\ref{clauses}.  

\begin{figure}
\vspace{1cm}
\includegraphics[width=0.4\textwidth]{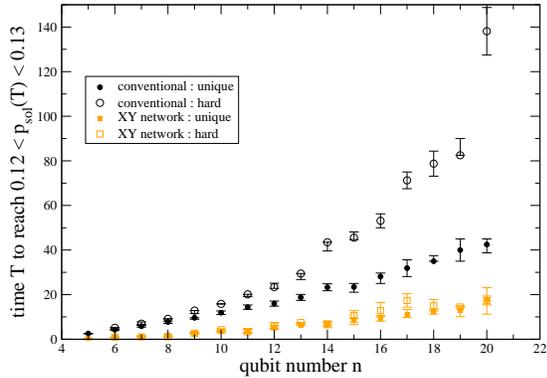}
\caption{Plot of the runtime necessary for an adiabatic evolution for 
the conventional scheme (\ref{conventional}) and the XY network in 
Eqs.~(\ref{triples}) and (\ref{xy}).
The hollow symbols correspond to especially hard problems within the 
class \mbox{\em exact cover-3}.}
\label{mintime} 
\end{figure}

Now, given the final problem Hamiltonian $H_{\rm out}$ above, what could be 
a suitable initial Hamiltonian $H_{\rm in}$? 
Obviously, $H_{\rm in}$ should not be diagonal in the 
$\sigma^z_\alpha$-basis -- in this case, we would have exact 
(instead of avoided) level crossings and an adiabatic evolution would be 
impossible. 
Therefore, most previous studies (see, e.g., \cite{Farhi}) adopted the 
form 
\bea
\label{conventional}
H_{\rm in}=\sum_\alpha L_\alpha\sigma^x_\alpha
\,.
\ea
On the other hand, inspired by the findings in Sec.~\ref{finite}, 
one could try to choose $H_{\rm in}$ in a way such that the change
form $H_{\rm in}$ to $H_{\rm out}$ is similar to a second-order
phase transition. 
The observation that symmetry-breaking/restoring phase transitions are 
often (though not always) of second order indicates one idea for doing this.
An apparent symmetry of $H_{\rm out}$ is its $O(2)$ invariance under 
rotations around the $z$-axis generated by  
\bea
\Sigma^z=\sum\limits_{\alpha=1}^{n}\sigma^z_\alpha
\,.
\ea
Now, if we start with the ferromagnetic Hamiltonian 
\bea
\label{xy}
H_{\rm in}
=
-\sum\limits_{\alpha,\beta=1}^{n}
M_{\alpha,\beta}
(\sigma^x_\alpha\sigma^x_\beta+\sigma^y_\alpha\sigma^y_\beta)
\,,
\ea
which is also $O(2)$ invariant, but whose degenerated ground states 
break this symmetry, one would expect that the resulting adiabatic 
quantum algorithm is very similar to a second-order phase transition, 
where the system finds it easier to stay in the ground state. 
In this way, physics intuition motivates a new way of constructing
quantum algorithms \cite{1st-vs-2nd}. 
Indeed, numerical simulations suggest that the algorithm described 
in Eqs.~(\ref{triples}) and (\ref{xy}) is superior (in average) to
the conventional scheme (\ref{conventional}), cf.~Fig.~\ref{mintime}. 

\section{Decoherence}\label{decoherence}

One of the (original) motivations for adiabatic quantum computing was the 
robustness of the ground state against decay and dephasing errors. 
However, this does not imply that these algorithms are completely free of
decoherence:
A finite temperature environment may heat up the system and cause 
excitations, which would result in computational errors (since the 
ground state, not an excited state, encodes the solution to our problem). 
Furthermore, non-equilibrium phenomena may also cause excitations at
zero temperature due to the coupling to an environment.
Since the energy gap typically becomes very small at some point of the 
interpolation, the algorithm will be very vulnerable near this point. 
Of course, the same applies to the critical point in more general 
quantum phase transitions. 
For an adiabatic version \cite{Roland} of the Grover \cite{Grover}
quantum search algorithm -- which corresponds to a first-order transition 
-- we found that the induced excitation probability strongly depends on the 
spectral function $f_{\rm bath}(\omega)$ of the environment and scales as 
\cite{Decoherence-Grover}
\bea
P_{\rm error}\sim\frac{f_{\rm bath}(\Delta E_{\rm min})}{\Delta E_{\rm min}}
\,,
\ea
where $\Delta E_{\rm min}$ is the minimum energy gap.
Hence, if $f_{\rm bath}(\omega)$ vanishes fast enough in the infrared 
$\omega\to0$, the error rate can be kept under control.
However, repeating the same analysis for the Ising model 
\cite{Decoherence-Ising} and the sweep 
through the second-order phase transition, we find that the 
excitation probability increases with system size $n$, independently 
of $f_{\rm bath}(\omega)$.
This suggests that second-order transitions are more suitable for 
adiabatic quantum computing -- but, at the same time, 
more vulnerable to decoherence 
(which requires some error correction techniques). 


\acknowledgements

This work was supported by the Emmy-Noether Programme of the German 
Research Foundation (DFG) under grant \# SCHU~1557/1-2,3.

\end{document}